\documentclass[a4paper]{jpconf}
\usepackage{graphicx}
\usepackage{mathrsfs}
\usepackage{amsmath,amssymb}
\usepackage{cite}
\usepackage{booktabs}

\def\beq{\begin{equation}}
\def\eeq{\end{equation}}
\def\bea{\begin{eqnarray}}
\def\eea{\end{eqnarray}}
\def\nn{\nonumber}

\def\N{{\cal N}}
\def\Sch#1{$\mathfrak{s}(#1)$}
\def\Q{\mathscr{Q}}
\def\S{\mathscr{S}}
\def\X{\mathscr{X}}
\def\SSch#1{$\mathfrak{s}(1/#1)$}
\def\SS2#1{$\mathfrak{s}(2/#1)$}
\def\Exo{$ \hat{\mathfrak{s}}(2/2)\; $}

\def\Fg{\mathfrak{g}}
\def\deg#1{ \mbox{deg}\ #1}
\newtheorem{prop}{Proposition}

\begin{document}
\title{Lowest weight representations of super Schr\"odinger algebras in low dimensional spacetime}

\author{N Aizawa}

\address{Department of Mathematics and Information Sciences, Graduate School of Science, \\ Osaka Prefecture University,
Nakamozu Campus, Sakai, Osaka 599-8531, Japan}

\ead{aizawa@mi.s.osakafu-u.ac.jp}

\begin{abstract}
 We investigate the lowest weight representations of the super Schr\"odinger algebras 
introduced by Duval and Horv\'athy. This is done by the same procedure as the semisimple Lie algebras. 
Namely, all singular vectors within the Verma modules are constructed explicitly then 
irreducibility of the associated quotient modules is studied again by the use of singular vectors. 
We present the classification of irreducible Verma modules for the super Schr\"odinger 
algebras in $ (1+1) $ and $ (2+1)$ dimensional spacetime with $\N = 1, 2 $ extensions. 
\end{abstract}

\section{Introduction}

  Throughout the last four decades nonrelativistic conformal symmetries have been 
observed in wide area of physics ranging from condensed matter to high energy 
(see for example \cite{Nie,Hagen,Henkel,Dunne,Sachdev,DuHo,HoMaSt,DuGiHo,MSW,StZak,LSZ,LSZ2,NS,Son,BaMcG,BG,ADV,MT} and references therein). 
In many cases the symmetries are described by the groups extending the Galilei symmetry group of nonrelativistic systems \cite{HaPle,NdelOR}. 
Those extended Galilei groups are non-semisimple as the original Galilei group. 
This may be an obstacle to develop systematic study of representations for such groups. 

  The simplest conformal extension of the Galilei group may be the Schr\"odinger groups \cite{Nie,Hagen} 
introduced in the studies of symmetry for free Schr\"odinger equation. Geometrical interpretation of the 
Schr\"odinger group revealed that the AdS/CFT correspondence has its nonrelativistic analogue \cite{Son,BaMcG} 
(see also \cite{DuGiHo}). 
Supersymmetric extensions of the Schr\"odinger group and 
its Lie algebra have also been discussed in connection with various physical systems 
such as fermionic oscillator \cite{BH,BDH}, spinning particles \cite{GGT}, 
nonrelativistic Chern-Simons matter \cite{LLM,DH,NSY}, Dirac monopole and magnetic vortex \cite{DH}, many-body 
quantum systems \cite{Gala}  
and so on. Some super Schr\"odinger algebras were constructed from the viewpoint of infinite dimensional 
Lie super algebra \cite{HU} or by embedding them to conformal superalgebras \cite{SY1, SY2}.

 Despite the numerous publications on the Schr\"odinger algebra and its supersymmetric extensions, 
the representation theories of them are not studied well. 
Especially there are a few works  based on representation theoretic viewpoint. 
We mention the followings: 
Projective representations of the Schr\"odinger group in 3 spatial dimension were constructed in \cite{Perroud}. 
Irreducible representations of the Schr\"odinger algebras up to 3 spatial dimension were investigated in 
\cite{DDM,Mur,FKS}. The author of \cite{Naka} studied the highest weight representations of 
the $ {\cal N} =  2 $ super Schr\"odinger algebra 
in 2 spatial dimension ("exotic" algebra in the terminology of \cite{DH}). 
To apply the (super) Schr\"odinger algebras for physical problems 
detailed study of irreducible representations of them will be helpful. 

  In the present work, we provide a classification of the Verma modules (a lowest weight representations) 
over the super Schr\"odinger algebras for simple cases, namely, one and two spatial dimensions with 
$ \N = 1, 2 $ supersymmetric extensions. The Schr\"odinger algebra discussed in this paper is centrally 
extended one. The central term corresponds to the mass of a system. 
In general, the Schr\"odinger algebra of fixed spacetime dimension has some distinct supersymmetric extensions 
even for fixed value of $\cal N.$ A systematic method to extend the Schr\"odinger algebra in $(n+1)$ dimensional 
spacetime to arbitrary $ \cal N $ has developed in \cite{DH}. 
We restrict our study of the representations to the super Schr\"odinger algebras introduced in \cite{DH}. 
We denote the centrally extended Schr\"odinger algebra in $(n+1)$ dimensional 
spacetime by \Sch{n}. The $ {\cal N} = 1, 2 $ extension of \Sch{1} are denoted by \SSch{1} and 
\SSch{2}, respectively. The $ \N=1 $ extension of \Sch{2} is also denoted by \SS2{1}. On the other hand there 
exist two $ \N=2 $ extensions for \Sch{2}. We denote one of them so-called standard by \SS2{2} and the another one 
so-called exotic by $ \hat{\mathfrak{s}}(2/2). $  
The superalgebra $ \mathfrak{s}(n/2) $ corresponds to the  $ N_+ = 1,\; N_- = 0 $ extended algebra  
in the notation of \cite{DH}. 

  The plan of this paper is as follows. We present the definition and structure of the super Schr\"odinger 
algebras in the next section. After giving the Verma modules in section 3 we give all the singular vectors 
and list of irreducible lowest weight modules without detailed calculation. 
The detailed proof is found in \cite{Naru,Naru2}. The employed procedure in the references is 
the same one as \cite{DDM,Mur}. This shows that the standard method of representation theory for 
semisimple Lie algebras is able to apply the super Schr\"odinger algebras, too.

\section{Definition and structure of super Schr\"odinger algebras}

 The bosonic Schr\"odinger algebra in $(n+1)$ dimensional spacetime  \Sch{n} is 
generated by the following transformations:   time translation $(H)$,  space 
translations $(P_a)$,  Galilei boosts $(G_a)$, spatial rotations $(J_{ab})$, 
dilatation $(D)$, conformal transformation $(K)$ and the central element $(M).$  
Contrast to the relativistic conformal algebra, there exists only one conformal generator. 
In this section we give the definition of the super Schr\"odinger algebras studied in this paper and 
their triangular decomposition which will be the key to study representations.

\subsection{Algebras in $(1+1)$ dimensional spacetime}

  The bosonic sector has six generators. Their nonvanishing commutators are given by
\bea
        \begin{array}{llll}
          [H, D] = 2H, & \quad [H, K]=D, & \quad [D, K ]= 2K, & \quad [P, G] = M, \\[3pt]
          [H, G] = P,  & \quad [D, G] = G, & \quad [P, D] = P, & \quad [P, K] = G.
        \end{array}
        \label{S1bosonic}
\eea
The $ \N = 1 $ extension \SSch{1} has three fermionic generators: $ \Q, \S, \X. $ They satisfy the 
anti-commutation relations 
\bea
  \begin{array}{lll}
      \{ \Q, \Q \} = -2H, & \quad \{ \S, \S \} = -2K, & \quad \{ \X, \X \} = -M, \\[3pt]
      \{ \Q, \X \} = -P, & \quad \{ \S, \X \} = -G, & \quad \{ \Q, \S \} = -D.
  \end{array}
  \label{S1N1fermionic}
\eea
As seen from the relations none of them are nilpotent. Nonvanishing bosonic-fermionic relations are listed below:
\beq
  \begin{array}{llll}
    [\Q, D] = \Q, & \quad [\Q, K] = \S, &  \quad [D, \S] = \S, &  \quad [H, \S] = \Q, \\[3pt]
    [\Q, G] = \X, &  \quad [P, \S] = \X.
  \end{array}
  \label{S1N1BS}
\eeq

 The $ \N=2 $ extension \SSch{2} has six fermionic generators $ \Q_a, \S_a, \X_a \ (a = \pm) $ 
and one additional bosonic generator $R$ corresponding to the $R$-parity. 
The generator $R$ commutes with all bosonic elements of \SSch{2}. 
The nilpotent fermionic generators satisfy the anti-commutation relations
\beq
  \begin{array}{lll}
     \{ \Q_{\pm}, \Q_{\mp} \} = - 2H, & \quad \{ \S_{\pm}, \S_{\mp} \} = - 2K, & \quad \{ \X_{\pm}, \X_{\mp} \} = -M,
     \\[3pt]
     \{ \Q_{\pm}, \X_{\mp} \} = -P, & \quad \{ \S_{\pm}, \X_{\mp} \} = -G, &  \quad 
     \{ \Q_{\pm}, \S_{\mp} \} = - D \mp R.
  \end{array}
  \label{S1N2fermionic2}
\eeq
Bosonic-fermionic commutators are given by
\beq
  \begin{array}{llll}
    [ \Q_{\pm}, D ] = \Q_{\pm}, & \quad [ \Q_{\pm}, K ] = \S_{\pm}, &  \quad [D, \S_{\pm} ] = \S_{\pm}, 
    & \quad [ H, \S_{\pm} ] = \Q_{\pm} 
    \\[3pt]
    [ \Q_{\pm}, G ] = \X_{\pm}, &  \quad [ P, \S_{\pm} ] = \X_{\pm}, & \quad [ R, \mathscr{A}_{\pm}] 
    = \pm \mathscr{A}_{\pm},
  \end{array}
  \label{S1N2BS2}
\eeq
where $ \mathscr{A}_a = \Q_a, \S_a, \X_a. $  
  
\subsection{Algebras in $(2+1)$ dimensional spacetime}

 In this spacetime there exists one spatial rotation $J.$ The bosonic algebra \Sch{2} has 
nine generators. Their nonvanishing commutators  are given by
\beq
        \begin{array}{llll}
          [H, D] = 2H, &  \quad   [H, K]=D, &  \quad   [D, K ]= 2K, &  \quad  [P_{\pm}, G_{\mp}] = 2M, \\[3pt]
          [H, G_{\pm}] = P_{\pm},  &  \quad   [D, G_{\pm}] = G_{\pm}, &  \quad   [P_{\pm}, D] = P_{\pm}, 
            &  \quad   [P_{\pm}, K] = G_{\pm}, \\[3pt]
          [J, G_{\pm}] = \pm G_{\pm}, &  \quad  [J, P_{\pm}] = \pm P_{\pm}. & 
        \end{array}
        \label{S2bosonic}
\eeq
The $ \N=1 $ extension \SS2{1} has four fermionic elements $ \Q, \S, \X_{\pm} $ subject to the 
relations:
\bea
  \begin{array}{lll}
      \{ \Q, \Q \} = -2H, & \quad \{ \S, \S \} = -2K, & \quad \{ \X_{\pm}, \X_{\mp} \} = -2M, \\[3pt]
      \{ \Q, \X_{\pm} \} = -P_{\pm}, & \quad \{ \S, \X_{\pm} \} = -G_{\pm}, & \quad \{ \Q, \S \} = -D.
  \end{array}
  \label{S2N1fermionic2}
\eea
The generators $ \X_{\pm} $ are nilpotent, however, $ \Q, \S $ are not. 
Nonvanishing bosonic-fermionic commutators are given by
\beq
  \begin{array}{llll}
    [\Q, D] = \Q, & \quad [\Q, K] = \S, &  \quad [D, \S] = \S, &  \quad [H, \S] = \Q, \\[3pt]
    [\Q, G_{\pm}] = \X_{\pm}, &  \quad [P_{\pm}, \S] = \X_{\pm}, & \quad
    [ J, \X_{\pm}] = \pm \X_{\pm}.
  \end{array}
  \label{S2N1BS2}
\eeq

  For \Sch{2} there exist two distinct $ \N=2 $ extensions, called  \textit{standard} and \textit{exotic}. 
The standard extension \SS2{2} has eight fermionic generators $ \Q_{\pm}, \S_{\pm}, \X_{\pm \pm}, \X_{\pm \mp} $ 
and additional bosonic $R$-parity operator.  
The fermionic sector is defined by the relations: 
\beq
  \begin{array}{lll}
    \{ \Q_{\pm}, \Q_{\mp} \} = - 2H, & \quad \{ \S_{\pm}, \S_{\mp} \} = - 2K, & \quad \{ \X_{++}, \X_{--} \} = -2M,,
    \\[3pt]
    \{ \X_{+-}, \X_{-+} \} = -2M, & \quad \{ Q_{\pm}, \X_{\sigma \mp} \} = -P_{\sigma}, 
    &  \quad \{ \S_{\pm}, \X_{\sigma \mp} \} = - G_{\sigma},  
    \\[3pt]
    \{ \Q_{\pm}, \S_{\mp} \} = -D \mp R,
  \end{array}
  \label{S2N2fermionic2}
\eeq
where $ \sigma=+, -.  $ Nonvanishing bosonic-fermionic relations are listed below:
\beq
  \begin{array}{llll}
    [ \Q_{\pm}, D ] = \Q_{\pm}, & \quad [ \Q_{\pm}, K ] = \S_{\pm}, &  \quad [D, \S_{\pm} ] = \S_{\pm}, 
    & \quad [ H, \S_{\pm} ] = \Q_{\pm} 
    \\[3pt]
    [ G_{\pm}, \Q_{\pm} ] = -\X_{\pm\pm}, &  \quad [G_{\pm}, \Q_{\mp} ] = - \X_{\pm\mp}, & \quad
    [ P_{\pm}, \S_{\pm} ] = \X_{\pm\pm}, & \quad [P_{\pm}, \S_{\mp}] = \X_{\pm\mp},
    \\[3pt]
    [ R, \mathscr{A}_{\pm}] = \pm \mathscr{A}_{\pm}, & \quad [R, \X_{\sigma\pm} ] = \pm \X_{\sigma\pm}, 
    & \quad [J , \X_{\sigma\pm}] = \sigma \X_{\sigma\pm},
  \end{array}
  \label{S2N2BS2}
\eeq
where $ \mathscr{A} = \Q, \S. $ 

   The exotic extension \Exo has six  fermionic generators $ \Q_{\pm}, \S_{\pm}, \X_{\pm} $ 
and additional bosonic $R$-parity operator.  
The fermionic generators are subject to the relations:
\beq
  \begin{array}{lll}
     \{ \Q_{\pm}, \Q_{\mp} \} = - 2H, & \quad \{ \S_{\pm}, \S_{\mp} \} = - 2K, & \quad \{ \X_{\pm}, \X_{\mp} \} = -M,
     \\[3pt]
     \{ \Q_{\pm}, \X_{\mp} \} = - P_{\mp}, & \quad \{ \S_{\pm}, \X_{\mp} \} = - G_{\mp}, & \quad 
     \{ \Q_{\pm}, \S_{\mp} \} = -D \mp (J + R),  
  \end{array}
  \label{exotic-fermionic2}
\eeq
and those for the bosonic-fermionic sector are
\beq
  \begin{array}{llll}
    [ \Q_{\pm}, D ] = \Q_{\pm}, & \quad [ \Q_{\pm}, K ] = \S_{\pm}, &  \quad [D, \S_{\pm} ] = \S_{\pm}, 
    & \quad [ H, \S_{\pm} ] = \Q_{\pm} 
    \\[3pt]
    [\Q_{\pm}, G_{\pm}] = 2\X_{\pm}, & \quad  [P_{\pm}, \S_{\pm}] = 2 \X_{\pm}, & \quad 
    [J, \Q_{\pm} ] = \mp \Q_{\pm}, & \quad [J, \S_{\pm}]= \mp \S_{\pm},
    \\[3pt]
    [R, \mathscr{A}_{\pm}] = \pm 2\mathscr{A}_{\pm}, & & &
  \end{array}
  \label{exotic-BS2}
\eeq
where $ \mathscr{A} = \Q, \S, \X. $ 
We remark that the $R$-parity operator for $ \N=2 $ extensions also commute with all bosonic generators. 

\subsection{Triangular decomposition}

  The grading and triangular decomposition for \Sch{n} 
introduced in \cite{DDM} are easily extended to the super Schr\"odinger algebras 
discussed in this paper. 
We define the following degree for \SSch{1}:
\bea
  & & 
  \deg{K} = 2, \qquad \deg{G} = \deg{\S} = 1, \qquad \deg{D} = \deg{M} = \deg{\X} = 0, 
  \nn \\
  & & 
  \deg{P} = \deg{\Q} = -1, \qquad \deg{H} = -2. \label{degs11}
\eea
It is immediate to see that \SSch{1} is ${\mathbb Z}$-graded by this degree. 
According to the sign of the degree one may define the triangular decomposition of 
\SSch{1} as follow:
\bea
  \mathfrak{s}(1/1) &=& \mathfrak{s}(1/1)^+ \oplus \mathfrak{s}(1/1)^0 \oplus \mathfrak{s}(1/1)^- 
  \nn \\
  &=& \{\ K, \ G \ \} \oplus \{\ D,\ M, \ \X \ \} \oplus \{\ P,\ \Q \ \}.
  \label{s11Tri}
\eea
Appropriate definitions of degree for other algebras also enable us to 
define the triangular decomposition. Below we give the decomposition. 
The algebras \SSch{2} and \SS2{1} are $ {\mathbb Z}^2$-graded:
\[
   \begin{array}{ccc} \toprule
         \mbox{decomposition} & & \mbox{generator (degree)} \\ \midrule
        \mathfrak{s}(1/2)^+ & \quad &  K (2,0), \ G (1,0), \ \S_+ (1,1), \ \S_- (1,-1),\ \X_+ (0,1) \\[5pt]
        \mathfrak{s}(1/2)^0 & & D (0,0), \ R (0,0), \ M (0,0) \\[5pt]
        \mathfrak{s}(1/2)^- & & H (-2,0), \ P (-1,0), \ \Q_+ (-1,1), \ \Q_-(-1,-1), \  \X_- (0,-1) 
        \\ \midrule
        \mathfrak{s}(2/1)^+ & \quad &  K (2,0), \ G_+ (1,1), \ G_- (1,-1), \ \S (1,0), \ \X_+ (0,1) \\[5pt]
        \mathfrak{s}(2/1)^0 & & D (0,0), \ J (0, 0), \ M (0,0) \\[5pt]
        \mathfrak{s}(2/1)^- & &  H (-2,0), \ P_+ (-1,1), \ P_- (-1,-1), \ \Q (-1,0), \ \X_- (0,-1) 
        \\ \bottomrule
   \end{array} 
\]

\medskip\noindent
While the algebras \SS2{2} and \Exo are $ {\mathbb Z}^3$-graded:
\[
   \begin{array}{ccc} \toprule
         \mbox{decomposition} & & \mbox{generator (degree)} \\ \midrule
         \mathfrak{s}(2/2)^+ & \quad &   K (2,0,0), \ G_+ (1,1,0), \ G_- (1,-1,0), \ \S_+ (1,0,1), \ \\[5pt]
                             &       &   \S_- (1,0,-1), \ \X_{++} (0,1,1), \ \X_{+-} (0,1,-1) \\[5pt]
         \mathfrak{s}(1/2)^0 & \quad &  D (0,0,0), \ J (0, 0, 0), \ M (0,0,0), \ R (0,0,0) \\[5pt]
         \mathfrak{s}(1/2)^- & \quad &  H (-2,0,0), \ P_+ (-1,1,0), \ P_- (-1,-1,0), \ \Q_+ (-1,0,1) \\[5pt]
                             &       &  \Q_- (-1,0,-1),\ \X_{-+} (0,-1,1), \ \X_{--} (0,-1,-1) 
         \\ \midrule
         \hat{\mathfrak{s}}(2/2)^+ & \quad &  K (2,0,0), \ G_+ (1,1,0), \ G_- (1,-1,0), \\[5pt]
                             &       &  \S_+ (1,-1,2), \ \S_- (1,1,-2), \ \X_+ (0,0,2) \\[5pt]
         \hat{\mathfrak{s}}(1/2)^0 & \quad &  D (0,0, 0), \ J (0, 0, 0), \ M (0,0, 0), \ R (0,0,0) \\[5pt]
         \hat{\mathfrak{s}}(1/2)^- & \quad &  H (-2,0,0), \ P_+ (-1,1,0), \ P_- (-1,-1,0), \\[5pt]
                             &       &  \Q_+ (-1,-1,2), \ \Q_- (-1,1,-2), \ \X_- (0,0,-2)
         \\ \bottomrule
   \end{array}
\]

\medskip
  One can introduce an algebra anti-automorphism for the super Schr\"odinger algebras. 
It is defined for the bosonic generators of \SSch{{\cal N}} by
\beq
  \omega(P) =  G, \quad \omega(H) = K, \quad \omega(D) = D, \quad \omega(M) = M, 
  \quad \omega(R) = R,
  \label{omega1-even}
\eeq
and those for the algebras in $ (2+1)$ dimension by
\beq
  \omega(P_a) =  G_{-a}, \quad \omega(H) = K, \quad \omega(D) = D, \quad \omega(M) = M, 
  \quad \omega(J) = J, \quad \omega(R) = R.
  \label{omega1-even2}
\eeq
The mapping of the fermionic elements of \SSch{1} is given by
\beq
  \omega(\Q) =  \S, \qquad \omega(\X) =  \X.
  \label{omega1-odd1}
\eeq
All the fermionic elements of the other algebras are obey the same transformation rule together 
with all the signature of the suffices being changed, e.g., $ \omega(\X_{+-}) = \X_{-+}. $ 
The mapping is involutive: $ \omega^2 = id. $ 
Let $ \mathfrak{g} $ be one of the superalgebras considered in this paper and let us denote 
its triangular decomposition by $ \Fg = \Fg^+ \oplus \Fg^0 \oplus \Fg^-. $ 
Then one see that $ \omega(\Fg^+) = \Fg^-, \ \omega(\Fg^0) = \Fg^0. $

\section{Verma modules and their reducibility}

The algebras under consideration admit the triangular decomposition. 
This allows us to define the lowest weight modules for the algebras. 
The lowest weight vector $ v_0 $ is defined by
\beq
   X v_0 = 0 \quad {}^{\forall}X \in \Fg^-, \qquad X v_0 = \Lambda(X) v_0 \quad {}^{\forall}X \in \Fg^0,
   \label{LWVdef}
\eeq
where $ \Lambda(X) $ is an eigenvalue of $X.$ 
The Verma module over $ \Fg $ is defined by
\beq
  V^{\Lambda} = U(\Fg^+) v_0, \label{VMdef}
\eeq
where $ U(\Fg^+)$ is the enveloping algebra of $ \Fg^+. $ 
Reducibility of the Verma modules may be investigated by singular vectors 
as in the cases of semisimple Lie algebras and the bosonic Schr\"odinger algebras \cite{DDM,Mur}. 
A singular vector $ v_s $ is defined as a homogeneous element of $V^{\Lambda}$ such that 
$ v_s \neq {\mathbb C} v_0 $ and 
$   \Fg^- v_s = 0. $ 
In this section we provide explicit formulae of the singular vectors and 
reducibility of the Verma modules for \textit{massive} case, that is, the generator $M$ has 
nonvanishing eigenvalues. 
Here we give only the outline of the calculation. 
Detailed proof  for $(1+1)$ dimensional algebras is found in \cite{Naru} and for 
$ (2+1)$ dimension will be published separately \cite{Naru2}. 

\subsection{$(1+1)$ Dimensional algebras}

 We start with \SSch{1}. 
 The lowest weight condition (\ref{LWVdef}) for \SSch{1} yields
\bea
 & & \Q \, v_0 = P \, v_0 = 0, \nn \\
 & & D v_0 = -d v_0, \quad M v_0 = m v_0, \quad  \X v_0 = \chi v_0,
 \label{LWVS11}
\eea
where $d \in {\mathbb R} $ is the conformal weight and the minus sign is for later 
convenience. The variable $\chi$ is of odd parity relating to the mass eigenvalue 
by the relation $ m=2 \chi^2. $ 
The Verma module (\ref{VMdef}) for \SSch{1} is given by
\beq
  V^d = \{  \ G^k K^{\ell} v_0, \ G^k K^{\ell} \S v_0 \ | \ k, \ell \in {\mathbb Z}_{\geq 0} \ \}.
  \label{VdS11}
\eeq
It is easy to see that the generator $D$ is diagonal on the basis given in (\ref{VdS11}). 
It follows that $ V^d $ is decomposed into a direct sum of the subspaces $ V_n^d $ spanned by the 
vectors $v$ satisfying $ D v = n v. $ 
The singular vector $ v_s $ belongs to $ V_n^d. $ One may set 
\beq
  v_s = \sum_k a_k v_k^{(n)}, \quad v_k^{(n)} \in V_n^d, \label{ansatzSV}
\eeq
and impose the conditions
\beq
   \Q v_s = P v_s = 0. \label{condSV}
\eeq
This yields some relations for $ a_k $ then by solving the relations one may determine the 
singular vectors up to overall constant. 

\begin{prop} \label{prop1}
  The Verma module $ V^d $ over \SSch{1} has precisely one singular vector iff $ d + 1/2 \in {\mathbb Z}_{\geq 0} $ 
  and it is given by
  \beq
     v_s = (G^2-2mK)^{d+1/2} (G-2\chi \S) v_0.   \label{SVN1}
  \eeq
\end{prop}

When the Verma module has singular vectors one find invariant subspaces constructed on the 
singular vector $ {\cal I} = U(\Fg^+) v_s. $ Thus the module is reducible. To find irreducible 
modules one may consider the quotient module $ V^d/{\cal I} $ then seeks singular vectors in 
the quotient module. 
In the present case it turns out that there is no singular vector in the quotient module. 
Therefore $ V^d/{\cal I} $ is irreducible. 

\begin{prop}\label{prop2}
 All irreducible lowest weight modules over \SSch{1} are listed as follows:
 \begin{enumerate}
    \item The Verma module $ V^d $ for $ d + 1/2 \notin {\mathbb Z}_{\geq 0} $
    \item The quotient module $ V^d/{\cal I} $ for $ d + 1/2 \in {\mathbb Z}_{\geq 0}$ 
 \end{enumerate}
 All irreducible modules given are infinite dimensional. 
\end{prop}

  We now turn to the superalgebra \SSch{2}. 
The lowest weight vector $ v_0 $ is defined by
\bea
  & & \Q_{\pm} v_0 = P v_0 = \X_- v_0 = 0,
  \label{LWS2} \\
  & & D v_0 = -d v_0, \quad M v_0 = m v_0, \quad R v_0 = r v_0.
  \nn
\eea
The Verma modules over \SSch{2} are given by
\beq
  V^{d,r} = \{ \ G^k K^{\ell} \S_+^a \S_-^b \X_+^c v_0 \ | \ k, \ell \in {\mathbb Z}_{\geq 0},\ a, b, c \in \{ 0, 1 \} \ \}.
  \label{VMS2}
\eeq
The same procedure as \SSch{1} leads us to the following propositions:
\begin{prop} \label{prop3}
  The Verma module $ V^{d,r} $ over \SSch{2} has precisely one singular vector iff $ d - 1/2 \in {\mathbb Z}_+ $ 
  and it is given by
    \bea
        & &  v_s^p = (G^2-2mK)^{d-1/2} u_0, \nn \\
        & & u_0 = (G \S_- \X_+ + m \S_+ \S_- + 2mK) v_0 + \frac{d+r+1}{2d+1} (G^2 - 2mK) v_0.
        \label{SVN2}
    \eea
\end{prop}
\begin{prop} \label{prop4}
 All irreducible lowest weight modules over \SSch{2} are listed as follows:
 \begin{enumerate}
    \item The Verma module $ V^{d,r} $ for $ d - 1/2 \notin {\mathbb Z}_+ $
    \item The quotient module $ V^{d,r}/{\cal I} $ for $ d - 1/2 \in {\mathbb Z}_+$ 
 \end{enumerate}
 where $ {\cal I} \subset V^{d,r} $ is the invariant submodule constructed on the singular vector (\ref{SVN2}): 
 $ {\cal I} = U(\mathfrak{s}(1/2)^+) v_s. $ 
 All irreducible modules given are infinite dimensional. 
\end{prop}

\subsection{$(2+1)$ Dimensional algebras}

  More involved in this case, however, one may use the  same procedure as $(1+1)$ dimensional algebras 
to study the reducibility of the Verma modules. 

We start with the superalgebra \SS2{1}. 
The lowest weight for \SS2{1} is defined by 
  \bea
    & &  \Q v_0 = P_{\pm} v_0 = \X_- v_0 = 0, \nn \\ 
    & &  D v_0 = -d v_0, \quad J v_0 = -j v_0, \quad M v_0 = m v_0. \label{LWS21}
  \eea
The Verma modules are constructed on the lowest weight vector:
\beq
 V^{d,j} = \{ \ G_+^k G_-^h K^{\ell} \S^a \X_+^b v_0 \ | \ k, h, \ell \in {\mathbb Z}_{\geq 0}, \ a, b \in \{ 0, 1 \} \ \}.
 \label{VMS21}
\eeq
The Verma modules contain a singular vector for discrete values of $d:$ 
\begin{prop} \label{prop5}
  The Verma module $ V^{d,j} $ over \SS2{1} has precisely one singular vector iff $ d +1  \in {\mathbb Z}_+ $ 
  and it is given by
  \beq
    v_s = (G_+ G_- - 2mK)^{d+1} (G_- \X_+ - 2m \S) v_0. \label{SV21}
  \eeq
\end{prop}

It turns out that the factor modules by the invariant submodule constructed on $ v_s$ do 
not contain any singular vectors. We thus have a classification of irreducible modules over \SS2{1}.
\begin{prop} \label{prop6}
 All irreducible lowest weight modules over \SS2{1} are listed as follows:
 \begin{enumerate}
    \item The Verma module $ V^{d,j} $ for $ d +1 \notin {\mathbb Z}_+ $
    \item The quotient module $ V^{d,j}/{\cal I} $ for $ d +1 \in {\mathbb Z}_+$ 
 \end{enumerate}
 where $ {\cal I} \subset V^{d,r} $ is the invariant submodule constructed on the singular vector (\ref{SV21}): 
 $ {\cal I} = U(\mathfrak{s}(2/1)^+) v_s. $ 
 All irreducible modules given are infinite dimensional. 
\end{prop}


Next we investigate the standard $\N=2$ extension \SS2{2}. 
The lowest weight vector for \SS2{2} is defined by
  \bea
     & & \Q_{\pm} v_0 = P_{\pm} v_0 = \X_{-+} v_0 = \X_{--}v_0 = 0, \nn \\
     & & D v_0 = -d v_0, \quad J v_0 = -j v_0, \quad R v_0 = r v_0, \quad M v_0 = m v_0, \label{LWS22}
  \eea
and application of a element of $ U(\mathfrak{s}(2/2)^+) $ on $ v_0 $ generates a basis of a  
Verma module over \SS2{1}:
  \beq
         V^{d,j,r} = \{ \ G_+^k G_-^h K^{\ell} \S_+^{a_1} \S_-^{a_2} \X_{++}^{b_1} \X_{+-}^{b_2} v_0 \ | \  
           k, h, \ell \in {\mathbb Z}_{\geq 0}, \ a_i, b_i \in  \{ 0, 1 \}  \ \}.
         \label{VMS22}
  \eeq
Reducibility of the Verma modules is similar to the other cases. We can show the following 
proposition on the existence of singular vectors. 
\begin{prop} \label{prop7}
  The Verma module $ V^{d,j,r} $ over \SS2{2} has precisely one singular vector iff $ d  \in {\mathbb Z}_+ $ 
  and it is given by
  \bea
   & & 
    v_s = (G_+ G_- - 2mK)^d (G_- \X_+ - 2m \S) u_0, \nn \\
   & & 
    u_0 = \{ G_-^2 \X_{++} \X_{+-} - 2m G_- (\S_+ \X_{+-} - \S_- \X_{++}) 
               + 4m^2 (\S_+ \S_- + K) \} v_0.
   \label{SVS22}
  \eea
\end{prop}

  Further search of singular vectors in the quotient modules conclude the following classification 
of irreducible modules:
\begin{prop} \label{prop8}
 All irreducible lowest weight modules over \SS2{2} are listed as follows:
 \begin{enumerate}
    \item The Verma module $ V^{d,j,r} $ for $ d  \notin {\mathbb Z}_+ $
    \item The quotient module $ V^{d,j,r}/{\cal I} $ for $ d \in {\mathbb Z}_+$ 
 \end{enumerate}
 where $ {\cal I} \subset V^{d,r} $ is the invariant submodule constructed on the singular vector (\ref{SVS22}): 
 $ {\cal I} = U(\mathfrak{s}(2/2)^+) v_s. $ 
 All irreducible modules given are infinite dimensional. 
\end{prop}

%
We now turn to investigation of the exotic $ \N=2 $ extension \Exo. 
The singular vectors and irreducible modules have been studied in \cite{Naka}. 
Here we give a closed form expression of the singular vectors and more precise 
classification of irreducible  modules. 
The lowest weight vector $v_0$ and the Verma modules $ V^{d,j,r}$ are defined as usual:
  \bea
     & & \Q_{\pm} v_0 = P_{\pm} v_0 = \X_- v_0 = 0, \nn \\
     & & D v_0 = -d v_0, \quad J v_0 = -j v_0, \quad R v_0 = 2r v_0, \quad M v_0 = m v_0. \label{LWexo}
  \eea
\beq
  V^{d,j,r} = \{ \ G_+^k G_-^h K^{\ell} \S_+^a \S_-^b \X_+^c v_0, \ | 
      \ k, h, \ell \in {\mathbb Z}_{\geq 0}, \ a, b, c \in \{ 0, 1 \} \ \}.
  \label{VMexo}
\eeq
The classification of irreducible modules is summarized in the following two 
propositions. 
\begin{prop} \label{prop9}
  The Verma module $ V^{d,j,r} $ over \Exo has precisely one singular vector iff $ d  \in {\mathbb Z}_+ $ 
  and $ r =-(d-j+2)/2. $ It  is given by
  \bea
   & & 
     v_s = (G_+ G_- - 2mK)^d u_0, \nn \\
   & & 
    u_0 =  (G_- \S_- \X_+ + M \S_+ \S_- + 2m K) v_0.
   \label{SVexo}
  \eea
\end{prop}

\begin{prop} \label{prop10}
 All irreducible lowest weight modules over \Exo are listed as follows:
 \begin{enumerate}
    \item The Verma module $ V^{d,j,r} $ if $ d  \notin {\mathbb Z}_+ $ nor $ r \neq -(d-j+2)/2 $
    \item The quotient module $ V^{d,j,r}/{\cal I} $ if $ d \in {\mathbb Z}_+$  and $ r =-(d-j+2)/2 $
 \end{enumerate}
  where $ {\cal I} \subset V^{d,r} $ is the invariant submodule constructed on the singular vector (\ref{SVexo}): 
 $ {\cal I} = U(\hat{\mathfrak{s}}(2/2)^+) v_s. $ 
 All irreducible modules given are infinite dimensional. 
\end{prop}

\section{Concluding remarks}

  We investigate the lowest weight representations of the 
super Schr\"odinger algebras in $ (1+1)$ and $(2+1)$ dimensional spacetime with 
$ \N = 1, 2 $ extensions. Our main result is a classification of 
irreducible lowest weight modules for nonvanishing mass.  This was done by the use of singular vectors. 
All irreducible modules are infinite dimensional. Finite dimensional irreducible  modules appear 
when the mass is set equal to zero \cite{Naru}. 
One may introduce a bilinear form analogous to the  Shapovalov form \cite{Sha} of the semisimple Lie algebra 
to the Verma modules. Let $ \mathfrak{g} $ be a super Schr\"odinger algebra discussed in this paper and 
$ V$ be a Verma module over it. 
We define the bilinear form $ (\ ,\ ) : V \otimes V \rightarrow {\mathbb C} $  
by the relations:
\beq
  (X v_0, Y v_0) = (v_0, \omega(X) Y v_0), \qquad (v_0, v_0) = 1, \quad X, Y \in U({\mathfrak g})
  \label{ShaForm}
\eeq
By the bilinear form one may define unitary representations of the super Schr\"odinger algebras. 
It is also easy to see that if  $ v_m, v_n \in V $ have different weight of the generator $D$, 
then they are orthogonal with respect to this form:
\beq
  (v_m, v_n) = 0.  \label{Ortho}
\eeq 
It follows that a singular vector of ${\mathfrak g}$ is orthogonal to any other vectors in $V.$ 
This is the same property as semisimple case. 
Thus one may analyze the reducibility of the Verma modules also via the bilinear form. 

  One may use the singular vectors to obtain invariant partial differential equations. 
This is an analogy to the cases of semisimple Lie algebras \cite{Kos,Dob88} and the bosonic Schr\"odinger 
algebras \cite{DDM, Mur,ADD,ADDS}. In the latter case the obtained partial differential equations are a hierarchy of 
free Schr\"odinger equations. 
To obtain the invariant equations we need a vector field representation of the algebra. 
Vector field representations for \SSch{1} and \SSch{2} are found in \cite{Naru}. 
Another link of the present work to physical problem is a a supersymmetric extension of the group theoretical 
approach to nonrelativistic holography discussed in \cite{AV}. It may also require vector field representations 
of the super Schr\"odinger algebras. 

  We restrict ourselves to the super Schr\"odinger algebras in this work. 
However there exist other algebras which may be regarded as nonrelativistic analogue of 
conformal algebras \cite{HaPle,NdelOR} (see also \cite{DuHo}) and their supersymmetric extensions \cite{Luki}. 
Physical importance of such algebras has been recognized widely \cite{DuHo,HoMaSt,StZak,LSZ,LSZ2,MT}, however, 
representation theory of such algebras are  far from completion.  
It is an important issue to develop representation theory for the algebras along the line of the present work. 
We announce that it has been done for one of the so-called conformal Galilei algebras in \cite{NaPh}.

\bigskip

\end{document}